\documentclass[prd,floatfix,preprintnumbers]{revtex4}
\usepackage{graphicx}
\usepackage{dcolumn}
\usepackage{physics}
\usepackage[lofdepth,lotdepth]{subfig}
\usepackage{float}

\newcommand {\ga} {\ {\raise-.5ex\hbox{$\buildrel>\over\sim$}}\ }
\newcommand {\la} {\ {\raise-.5ex\hbox{$\buildrel<\over\sim$}}\ }
\DeclareMathOperator{\sgn}{sgn}

\def\be{\begin{equation}}
	\def\ee{\end{equation}}
\def\ba{\begin{eqnarray}}
	\def\ea{\end{eqnarray}}

\begin{document}
	\title{Observational constraints on inflection point quintessence with a cubic potential}
	\author{S. David Storm and Robert J. Scherrer}
	\affiliation{Department of Physics and Astronomy, Vanderbilt University,
		Nashville, TN  ~~37235}
	
	\begin{abstract}
	We examine the simplest
	inflection point quintessence model, with a potential
	given by $V(\phi) = V_0 + V_3 \phi^3$.  This model
	can produce either asymptotic de Sitter expansion
	or transient acceleration, and we show that it
	does not correspond to either pure freezing or thawing
	behavior.
	We derive observational
	constraints on the initial value of the scalar field, $\phi_i$,
	and $V_3/V_0$ and find that small values of
	either $\phi_i$ or $V_3/V_0$ are favored.  While most of the observationally-allowed
	parameter space yields asymptotic de Sitter evolution,
	there is a small region, corresponding
	to large $V_3/V_0$ and small $\phi_i$, for which the current
	accelerated expansion is transient.  The latter behavior is potentially
	consistent with a cyclic universe.
	\end{abstract}
	
	\maketitle
	
	\section{Introduction}
	According to observational data \cite{0804.4142,0901.4804,1004.1711,1105.3470,1212.5226,1303.5076,1401.4064}, the Universe is composed of approximately 70\% dark energy, which is a 
	negative-pressure component, and roughly 30\% nonrelativistic matter, which includes baryons
	and dark matter.  Although current observations are consistent with a cosmological constant and cold dark
	matter ($\Lambda$CDM), a dynamical equation of state cannot be ruled out.  One widely-studied
	set of dynamical models is quintessence, which employs
	a time-dependent scalar field, $\phi$, with an associated potential $V(\phi)$
	\cite{PUPT-1072,hep-th/9408025,astro-ph/9707286,gr-qc/9711068,astro-ph/9708069,astro-ph/9809272,astro-ph/9812313}.  (See Ref. \cite{hep-th/0603057} for a review).
	
	Dark energy is characterized by its
	equation of state parameter, $w$, the ratio of dark energy pressure to density:
	\be
	\label{w}
	w=p/\rho.
	\ee
	For a cosmological constant $\Lambda$, we have
	$\rho = constant$ and $w = -1$.
	For a scalar field $\phi$ evolving in a potential $V(\phi)$, the pressure and density that determine $w$ are given by
	\begin{equation}
	p_\phi = \frac{\dot{\phi}^2}{2} - V(\phi),
	\end{equation}
	and 
	\begin{equation}
	\label{rhophi}
	\rho_\phi = \frac{\dot{\phi}^2}{2} + V(\phi),
	\end{equation}
	where the dot denotes the time derivative throughout.
	Observations indicate that $w \approx -1$ today.
	In order to achieve this with quintessence,
	we require that $\dot \phi^2/2 \ll V(\phi)$.  One way to achieve this result is to begin with $\dot \phi \approx 0$ and take the present day to
	correspond to the time immediately after the field begins to roll downhill in the potential; in these ``thawing" models, $w$ is initially equal to $-1$ and increases slightly up to the present.  Alternately, one can allow the field to roll downhill with $w > -1$,
	but then to reach a sufficiently flat region of the potential that $w$ approaches $-1$ asymptotically.  Such models have been dubbed ``freezing" models.  (See Ref. \cite{astro-ph/0505494} for a discussion).  We will see that inflection point quintessence,
	proposed in Ref. \cite{1306.4662}, does not fit neatly into either category but
	instead corresponds to a hybrid of the two.
	
	Scalar fields with an inflection point in the potential were first proposed as models for inflation \cite{hep-ph/0605035,hep-ph/0610134,hep-ph/0608138,0705.3837,0707.2848,0708.2798,0705.4682,0806.4557,0810.4299,1004.3724,1101.6046,1103.5758,1203.6892,1211.1707,1305.6398}.	
	The simplest such models correspond to a cubic potential with
	an inflection point at $\phi = \phi_0$:
	\begin{equation}
	V(\phi) = V_0 + V_3 (\phi-\phi_0)^3,
	\end{equation}
	where $V_0$ and $V_3$ are constants.  As noted
	in Ref. \cite{0708.2798}, this model arises naturally
	in the context of string theory.  Note that $\phi$ can be translated by a constant without changing any physically-observable quantities, so it
	is simplest to take the inflection point to be at $\phi=0$,
	giving
	\begin{equation}
	\label{cubic}
	V(\phi) = V_0 + V_3 \phi^3.
	\end{equation}
	In both inflation and quintessence, a value
	of $w \approx -1$ is achieved as $\phi$ reaches the inflection point.
	
	An interesting feature of the evolution of $\phi$ in this potential is
	that the late-time behavior of $\phi$ can vary widely depending on the values
	of $V_0$, $V_3$, and $\phi_i$ (the initial value of $\phi$).  For some values of
	these parameters, $\phi$ evolves smoothly to zero as $t \rightarrow \infty$,
	yielding an asymptotic de Sitter evolution.  However, for other values, $\phi$
	evolves through the inflection point at $\phi = 0$, corresponding to a transient period of
	acceleration.  The latter possibility is particularly interesting because an eternally accelerating
	universe presents a problem for string theory, inasmuch as the S-matrix in this case is ill-defined
	\cite{Hellerman:2001yi,Fischler:2001yj}.  Consequently, a great deal of effort has gone into the development of models in which the
	observed acceleration is a transient phenomenon \cite{Barrow:2000nc,Cline:2001nq,Cardenas:2002np,Sahni:2002dx,Blais:2004vt,Bilic:2005sp,GCG,Carvalho:2006fy,
		Bento:2008yx,Cui:2013tna}.  When $\phi$ evolves past $0$,
	our model represents another example of this
	sort of transient acceleration.
	
	The goal of this investigation is to first determine the model parameters that
	are allowed by present-day observations, and then to investigate whether both types of future evolution (de Sitter versus transient acceleration) are consistent with the allowed parameter range.  In the next section, we discuss
	the general features of the evolution of $\phi$ in this model.  In Sec. III, we derive observational limits on the model parameters and explore whether these
	limits are consistent with both types of asymptotic evolution for $\phi$.  Our
	main results are summarized in Sec. IV.
	
	\section{The evolution of inflection point quintessence}
	
	The equation of motion for a quintessence scalar field is 
	\begin{equation}
		\label{phievol}
		\ddot{\phi} + 3H\dot{\phi} + \frac{dV}{d{\phi}} = 0,
	\end{equation}
	where the Hubble parameter $H$ is 
	\begin{equation}
		H^2 \equiv \Bigr(\frac{\dot{a}}{a}\Bigr)^2 = \frac{\rho_\phi + \rho_M}{3}.
	\end{equation}
	Here, $a$ is the scale factor, which we normalize to $a=1$ at the present, $\rho_\phi$ is the scalar field energy density given by Eq. (\ref{rhophi}), and $\rho_M$ is
	the matter density.
	We take $\hbar = c = 8 \pi G = 1$ throughout.
    The matter density is given by
	\begin{equation}
		\rho_M = \rho_{M0}a^{-3},
	\end{equation}
	where $\rho_{M0}$ is the present-day matter density.
	We use the cubic potential given by Eq. (\ref{cubic}) for $V(\phi)$.  While
	other potentials can correspond to inflection point quintessence (see Sec. IV), the cubic potential is the simplest, and it has the interesting
	property of yielding both asymptotic de Sitter or transient acceleration depending on the model parameters.
	
	In constraining quintessence models, the observable quantities of interest depend only on the evolution of $w$ as a function of $\Omega_\phi \equiv \rho_\phi/(\rho_\phi + \rho_M)$.  This evolution is given by \cite{0712.3450}
	\begin{equation}
	\frac{dw}{d \Omega_\phi} = \frac{3(1+w)(1-w)+\lambda(1-w)\sqrt{3(1+w)\Omega_\phi}}
	{3w\Omega_\phi(1-\Omega_\phi)},
	\end{equation}
	with $\lambda \equiv (1/V)(dV/d\phi)$.  Hence, the observable quantities depend on the potential only through
	$\lambda$.  In our cubic inflection point model,
	\begin{equation}
	\lambda = \frac{3 \phi^2}{(V_0/V_3) + \phi^3}. 
	\end{equation}
	Therefore, the observable quantities in this model depend only on $V_3/V_0$, rather than on $V_3$ and $V_0$ independently.
	
	We will assume that Hubble friction drives $\dot \phi$ to zero in the early
	universe, so that $\phi$ begins initially at some value $\phi_i$ with
	$\dot \phi_i = 0$.  With these assumptions, the cubic inflection point quintessence model has only two free parameters:  $V_3/V_0$ and $\phi_i$.
	For a fixed value of $V_3/V_0$, the asymptotic evolution of $\phi$ depends
	on the value of $\phi_i$: for a sufficiently large value of $\phi_i$, the
	scalar field evolves through the inflection point at $\phi=0$, while for smaller values
	of $\phi_i$, the field evolves asymptotically to $\phi =0$ as $t \rightarrow
	\infty$.  However, there exists a critical value of $V_3/V_0 \approx 0.77$;
	for values of $V_3/V_0$ below this value,
	$\phi$ always evolves asymptotically to zero regardless
	of the value of $\phi_i$. This behavior was first noted
	for inflation by Itzhaki and Kovetz \cite{0810.4299} and later for
	quintessence by Chang and Scherrer \cite{1306.4662}.

	\section{Observational limits}
	
	To derive observational limits on inflection point quintessence, we numerically
	integrated the evolution of $\phi$ for a range
	of values of $V_3/V_0$ and $\phi_i$ using
	CAMB \cite{Lewis:1999bs,Howlett:2012mh}. Consistency with observational data (the temperature, polarization, and lensing of the Planck 2018 likelihood data \cite{Aghanim:2019ame,Aghanim:2018oex}, baryon acoustic oscillations (BAO) data from 6dF, SDSS DR7, and SDSS DR12 galaxy surveys \cite{Beutler:2012px,Ross:2014qpa,Alam:2016hwk}, and the Type Ia Supernovae Pantheon data \cite{Scolnic:2017caz}) was
	determined using 
	Cobaya \cite{2005.05290} along with a Monte Carlo Markov Chain sampler \cite{Lewis:2002ah,Lewis:2013hha}, which used the fast-dragging procedure described in \cite{math/0502099}. 
	
\begin{figure}[H]
		\centering
		\includegraphics[width=0.8\textwidth]{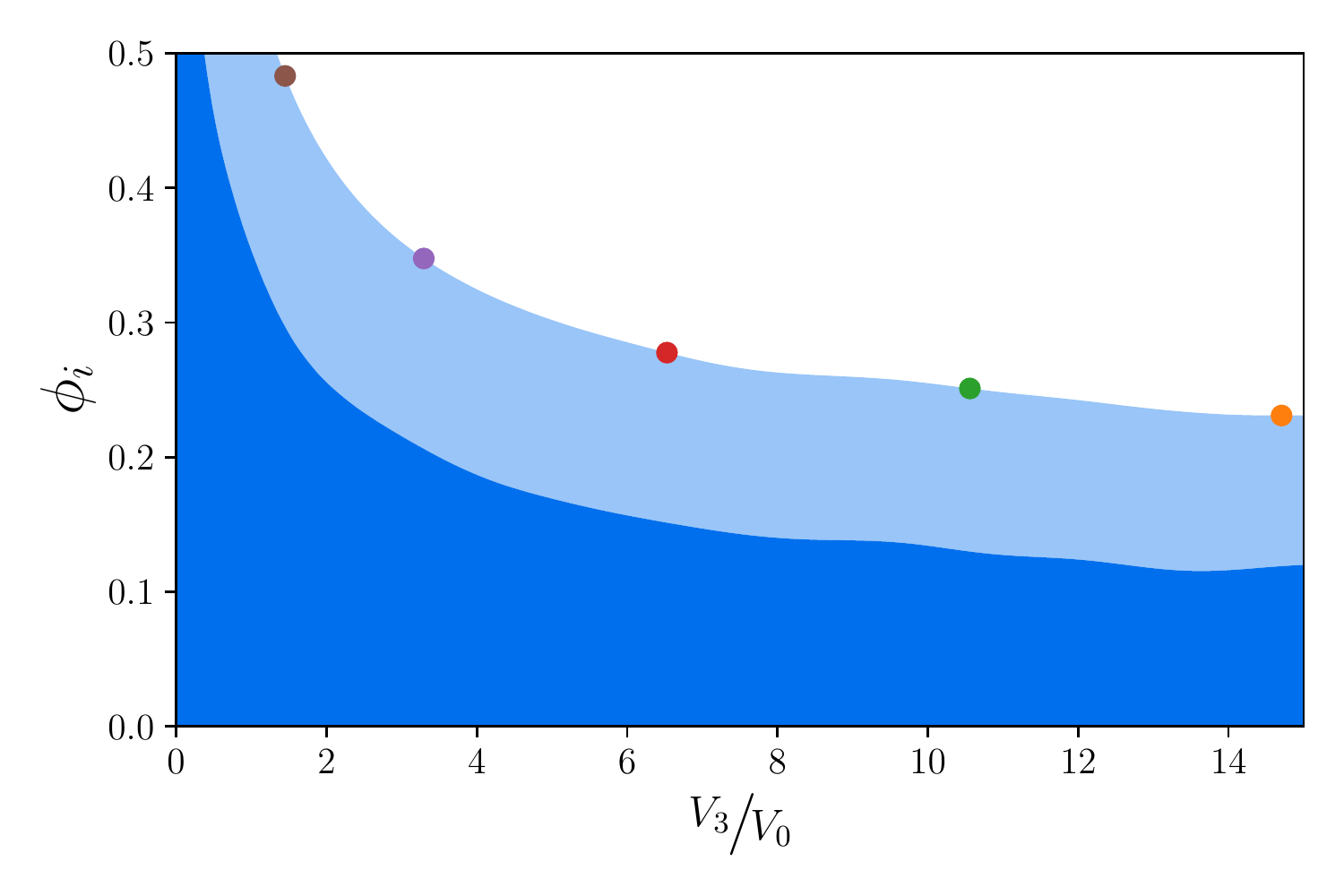}
		\caption{Planck 2018 CMB + SN + BAO $1\sigma$ (dark blue) and $2\sigma$ (light blue) contours for the inflection point quintessence model with
			the cubic potential of Eq. (\ref{cubic}), as a function of
			$V_3/V_0$ and $\phi_i$. The
			colored circles label the curves in Fig. 3 showing the evolution
			of $w$ for the corresponding values of $V_3/V_0$ and $\phi_i$.}
		\label{fig:main_plots}
	\end{figure}

		Our main result is shown in Fig. \ref{fig:main_plots}, which depicts the observational constraints on our model parameters.  For fixed $\phi_i$,
		the observations favor smaller values of $V_3/V_0$ and vice-versa.  This
		makes sense, since smaller values of $V_3/V_0$ and $\phi_i$ drive the model
		to more closely resemble the observationally-favored $\Lambda$CDM model.
		Indeed, as $\phi_i\rightarrow 0$ for fixed $V_3/V_0$ or $V_3/V_0 \rightarrow 0$ for fixed $\phi_i$, the model becomes indistinguishable from
		$\Lambda$CDM; this is reflected in the contours shown in Fig. \ref{fig:main_plots}.
		
\begin{figure}[H]
	\centering
	\includegraphics[width=1.\textwidth]{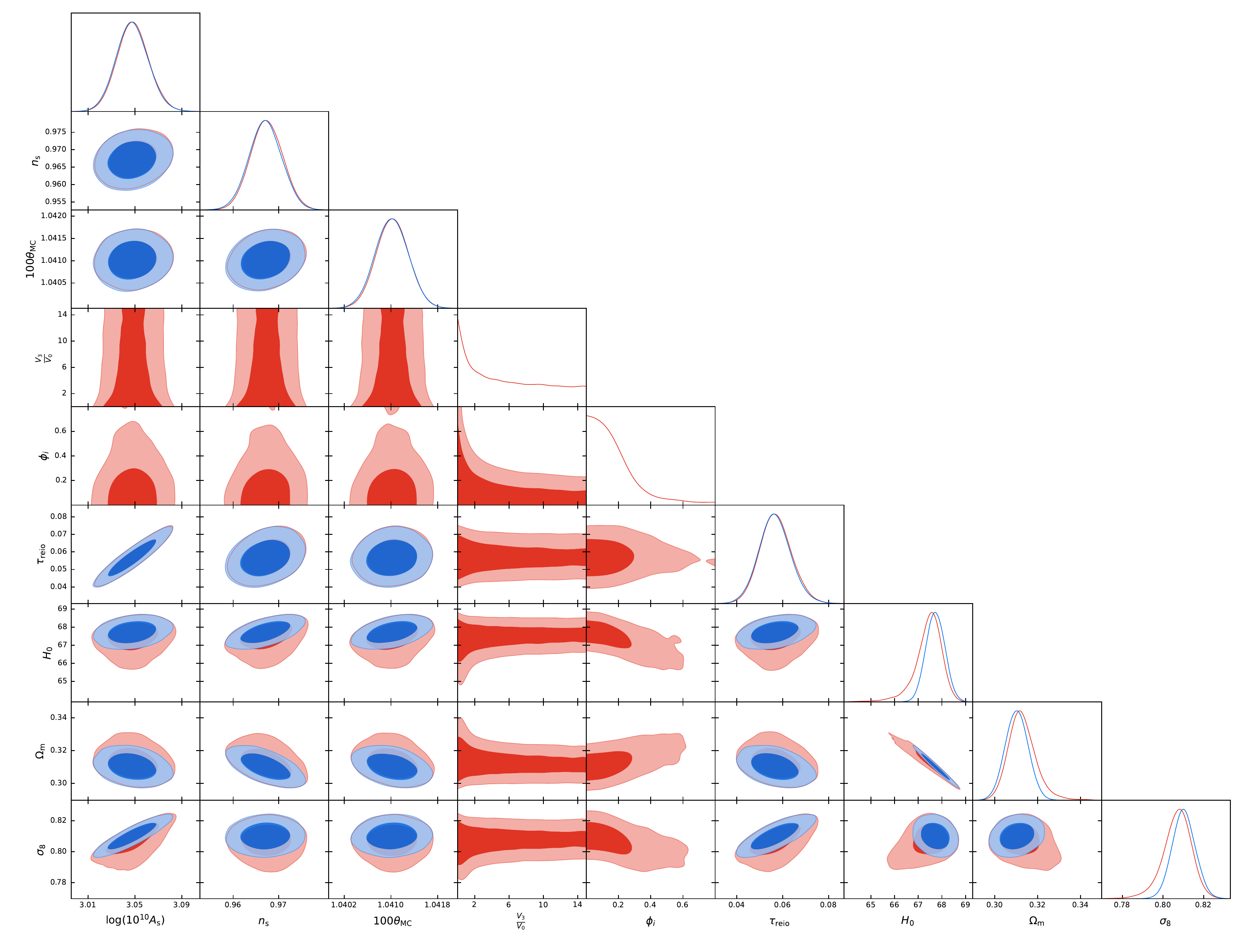}
	\caption{Planck 2018 CMB + SN + BAO $1\sigma$ (dark region) and $2\sigma$ (light region) contours for the $\Lambda$CDM model (blue) and the inflection point quintessence model (red) with
		the cubic potential of Eq. (\ref{cubic}), as a function of
		$V_3/V_0$ and $\phi_i$.}
	\label{fig:parameters_triangle}
	\end{figure}

		Upper and lower limits at the 68\% and 95\% C.L. for various other cosmological parameters using the inflection point quintessence model are given in Table \ref{table:parameterValues}. The inflection point quintessence model is compared to the standard $\Lambda$CDM model in Fig. \ref{fig:parameters_triangle}, which also includes the $\chi^2$ values.

\begin{table}
	\centering
	\begin{tabular} { l  c| l  c}
		
		Parameter &  68\% limits &  95\% limits\\
		\hline
		{$\log(10^{10} A_\mathrm{s})$} & $3.048\pm 0.014            $ & $3.048^{+0.028}_{-0.026}     $\\
		
		{$n_\mathrm{s}   $} & $0.9674\pm 0.0035          $ & $0.9674^{+0.0069}_{-0.0068}$\\ 
		
		{$100\theta_\mathrm{MC}$} & $1.04103\pm 0.00028        $ & $1.04103^{+0.00055}_{-0.00054}$\\
		
		{$\tau_\mathrm{reio}$} & $0.0570^{+0.0064}_{-0.0072}$ & $0.057^{+0.014}_{-0.013}   $\\
		
		$H_0                       $ & $67.42^{+0.63}_{-0.39}     $ & $67.4^{+1.1}_{-1.2}        $\\
		
		$\Omega_\mathrm{m}         $ & $0.3129^{+0.0052}_{-0.0070}$ & $0.313^{+0.013}_{-0.013}   $\\
		
		$\sigma_8                  $ & $0.8069^{+0.0075}_{-0.0059}$ & $0.807^{+0.014}_{-0.015}   $\\
		
		$\chi^2                    $ & $3832\pm 43                $ & $3832\,({\nu\rm{:}\,906.5})$\\
		\hline
	\end{tabular}
	
	\caption{Results for the marginalized means, 68\% limits, and 95\% limits for the initial super-horizon amplitude of curvature perturbations $A_s$, the scalar spectral index $n_s$, the approximation to the observed angular size of the sound horizon at recombination $\theta_{MC}$, the optical depth to reionization $\tau_{reio}$, the Hubble constant $H_0$, the matter density parameter $\Omega_m$, the late-time clustering amplitude $\sigma_8$, and the $\chi^2$ in the inflection point quintessence model.}\label{table:parameterValues}
\end{table}

\begin{figure}[htpb!]
	\centering
	\includegraphics[width=0.8\textwidth]{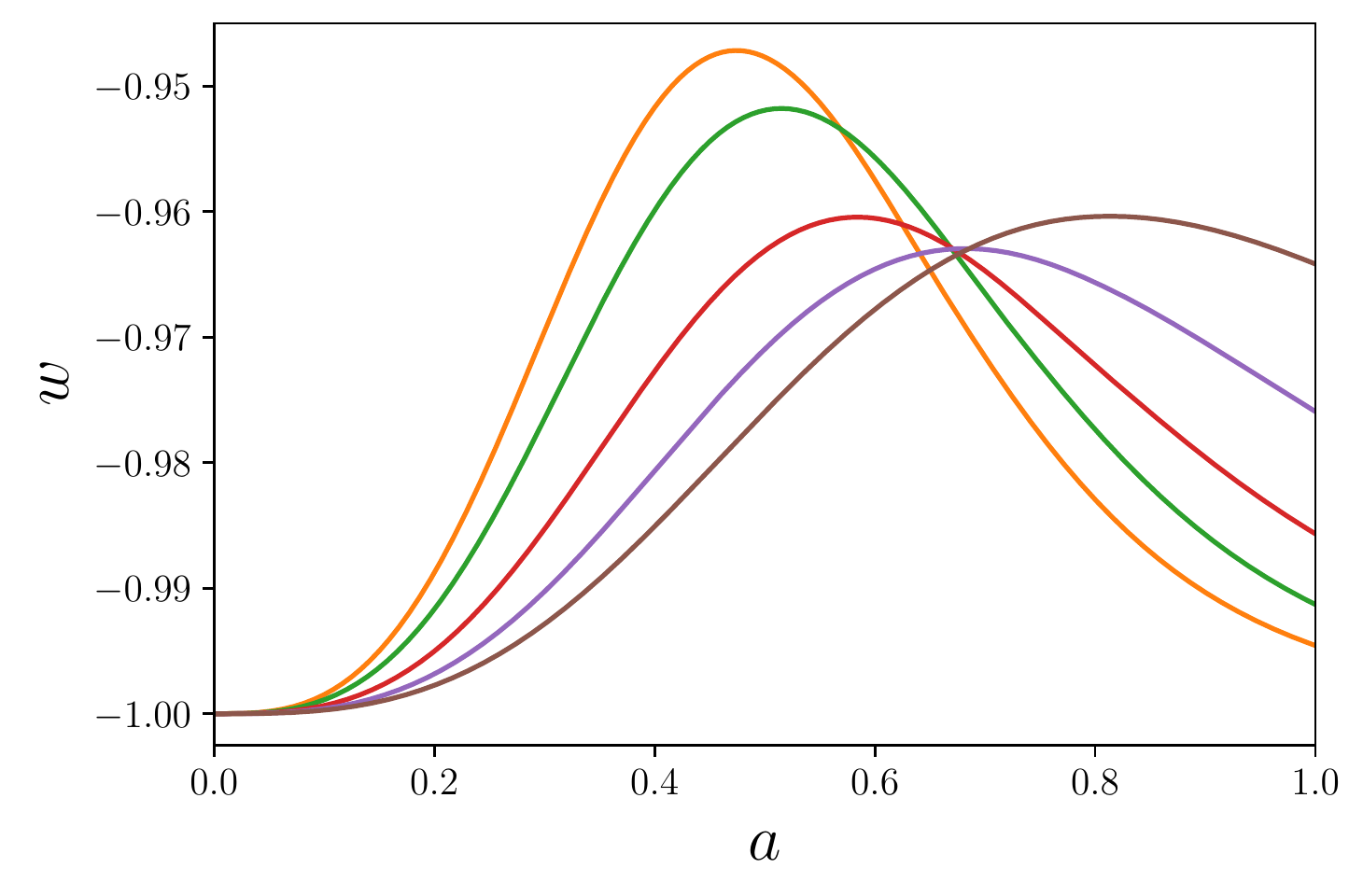}
	\caption{Evolution of the inflection point quintessence equation of state parameter $w$ as a function of the scale factor $a$, normalized to $a=1$ at present, for five representative values of the model parameters.  The color of each curve corresponds to the color of the corresponding point in parameter space indicated in Fig. \ref{fig:main_plots}.}
	\label{fig:globfig}
\end{figure}

		Fig. \ref{fig:globfig} depicts the evolution of $w$ as a function
		of the scale factor $a$ for different parameter values along the 2$\sigma$ contour.
		These curves illustrate the fact that inflection point
		quintessence cannot be treated as either a purely freezing or purely thawing model.  Instead, the scalar
		field begins with $w =-1$.  As the field rolls downhill,
		$w$ increases (thawing behavior) eventually reaching
		a maximum value.  Then as the field approaches the inflection point, $w$ decreases toward $-1$ (freezing behavior).
		
		These results further illustrate the extent to which current observations
		drive this model toward $\Lambda$CDM.  Even at the 2$\sigma$ level,
		the value of $w$ never increases beyond $-0.95$.
		These curves also illustrate the existence of
		a ``pivot redshift" $z_p =0.37$ noted by
		Alam et al. \cite{Alam:2016hwk}.  This redshift
		corresponds to $a_p = 0.73$.  The supernova data are
		most strongly constraining at this value of $a$,
		causing all of our curves to pass through the
		same value of $w$ near this point.

\begin{figure}[H]
	\centering
	\includegraphics[width=0.8\textwidth]{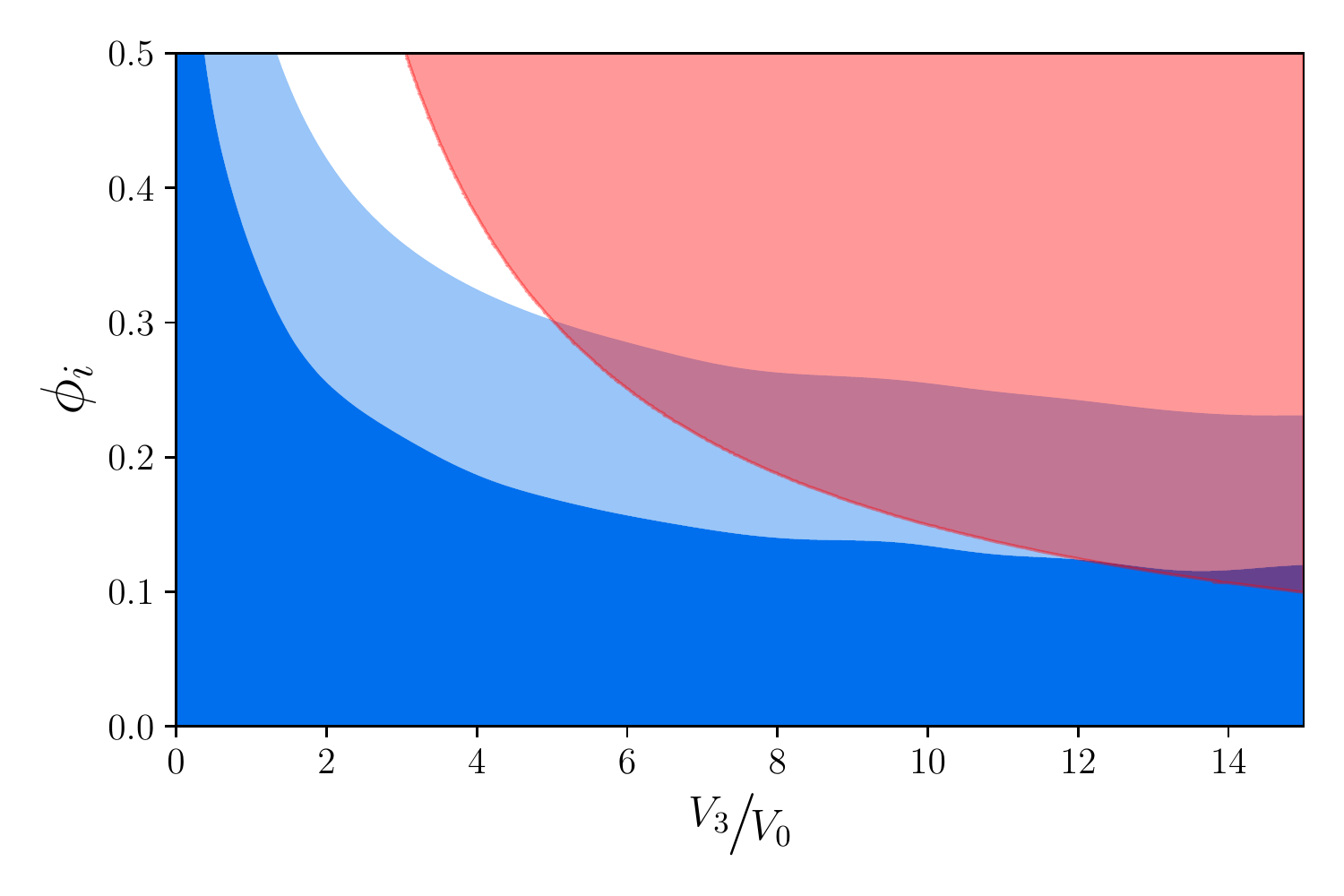}
	\caption{Planck 2018 CMB + SN + BAO $1\sigma$ (dark blue) and $2\sigma$ (light blue) contours for the inflection point quintessence model with the region (red) overlaid atop the parameter space where $\phi$ will eventually pass through the inflection point.}
	\label{fig:inflection_main}
\end{figure}

	In Fig. \ref{fig:inflection_main}, we address the key question of interest
for this model:  do current observations allow both types of asymptotic
behavior for the scalar field?  The red region in this figure corresponds
to the parameter values for which $\phi$ evolves through the inflection point, rather than evolving asymptotically to $\phi = 0$.  The observations
clearly favor the latter behavior; over most of the allowed parameter range,
$\phi \rightarrow 0$ asymptotically.  However, for large values of $V_3/V_0$,
there is an observationally-allowed range of values for $\phi_i$ for
which $\phi$ evolves through the inflection point.  Thus, the cubic
inflection point model can support a transient accelerated expansion that is
compatible with current observations.

\section{Results}

Our results indicate that the inflection point model with the cubic potential
given by Eq. (\ref{cubic}) can be made consistent with present observational constraints.  This model is poorly described
as either a thawing or freezing model; the equation of state parameter $w$
increases from $-1$ to a maximum value and then decreases back toward $-1$.
However, even for models that are nearly ruled out
at the $2\sigma$ level, $w$ never evolves above a value of $-0.95$, as current observations
continue to drive acceptable models closer to $\Lambda$CDM.  While most of
the observationally-allowed parameter space corresponds to a value of $\phi$ that
asymptotes to $\phi=0$, there is a small region of parameter space for which
$\phi$ can evolve through the inflection point, corresponding to the interesting case of transient
acceleration.

These two cases result in very different predictions for the future evolution of the Universe.  When $\phi$ evolves asymptotically to 0, the future evolution of the Universe becomes indistinguishable from $\Lambda$CDM.  Alternately, when $\phi$ evolves through the inflection point, the future evolution depends on the shape of the potential for $\phi < 0$.  If we simply extrapolate Eq. (\ref{cubic}) to negatives values of $\phi$, we see that $V(\phi)$ eventually becomes negative.  The evolution of the universe with negative quintessence potentials has long been studied \cite{Felder,Alam}.  When the total energy density (matter plus quintessence) reaches zero, the universe stops expanding and begins a contracting phase.  As noted in Refs. \cite{Turok,Ijjas,Andrei}, this provides a
natural way to construct a cyclic model for the Universe.

The cubic potential is by no means the only model that can
correspond to inflection point quintessence.  Any model of the form
\begin{eqnarray}
V(\phi) &=& V_0 + V_n \sgn(\phi)\phi^n,~~~n~{\rm even},\\
V(\phi) &=& V_0 + V_n \phi^n,~~~n~{\rm odd},
\end{eqnarray}
with $n \ge 2$
has an inflection point at $\phi=0$ and can therefore serve as
a model for inflection point quintessence.
While the cubic model is the simplest and most natural inflection point quintessence model, these other models
can also yield interesting behavior, resulting in either
asymptotic de Sitter evolution or transient acceleration,
depending on the value of $n$ and the particular model parameters \cite{1306.4662}.
We have not attempted a detailed analysis of these
other models, but we expect qualitatively similar results
for the allowed parameter ranges; current observations
will tend to favor small values of either  $\phi_i$
or $V_n/V_0$.

	\section*{Acknowledgments}
	
	R.J.S. was supported in part by the Department of Energy (DE-SC0019207). S.D.S. thanks Antony Lewis for helpful discussions.

\end{document}